\documentclass[pre,aps,twocolumn,showpacs,superscriptaddress]{revtex4-1}

        \newif\ifdraft \draftfalse  
        \newif\ifpreparepdf \preparepdftrue 
\drafttrue          

\usepackage[pdftex]{color,graphicx}
    \usepackage{tikz}
    \usepackage{graphicx,amssymb,amsbsy}
    \usepackage[normalem]{ulem}
    \usepackage{ifthen}
  \usepackage[pdftex,colorlinks]{hyperref} 

    \graphicspath{{../figs/}{../Fig/}}  

\ifpreparepdf 

    \newcommand{\arXiv}[1]{
              {\tt \href{http://arXiv.org/abs/#1}{arXiv:#1}}}
\else  

    \newcommand{\arXiv}[1]{ {\tt arXiv:#1}}
\fi

        \ifdraft 
   \newcommand{\PCedit}[1]{{\color{blue}#1}} 
   \newcommand{\PC}[2]{\begin{quote}\PCedit{[#1 Predrag] #2}\end{quote}}
   \newcommand{\PCcut}[1]{{\color{orange}\sout{#1}}} 
   \newcommand{\APWedit}[1]{{\color{blue}#1}} 
   \newcommand{\APWcut}[1]{{\color{orange}\sout{#1}}} 
   \newcommand{\APW}[2]{\begin{quote}\APWedit{[#1 Ash] #2}\end{quote}}
   \newcommand{\KYSedit}[1]{{\color{blue}#1}}
   \newcommand{\KYScut}[1]{{\color{orange}\sout{#1}}} 
   \newcommand{\KYS}[2]{\begin{quote}\KYSedit{[#1 Kimb] #2}\end{quote}}
   \newcommand{\ToDo}[1]{{\bf\color{red}#1}}
   \newcommand{\TWRUN}[2]{\ensuremath{\textrm{TW}_{#1}}}    
   \newcommand{\RPORUN}[2]{\ensuremath{\textrm{RPO}_{#1}}}  
        \else  
   \newcommand{\PC}[2]{}{}
   \newcommand{\PCedit}[1]{{#1}}
   \newcommand{\PCcut}[1]{}
   \newcommand{\APW}[2]{}{}
   \newcommand{\APWedit}[1]{{#1}}
   \newcommand{\APWcut}[1]{}
   \newcommand{\KYS}[2]{}{}
   \newcommand{\KYSedit}[1]{{#1}}
   \newcommand{\KYScut}[1]{}
   \newcommand{\ToDo}[1]{}
   \newcommand{\TWRUN}[2]{\ensuremath{\textrm{TW}_{#1}}}
   \newcommand{\RPORUN}[2]{\ensuremath{\textrm{RPO}_{#1}}}
        \fi

\newcommand{\rf}     [1] {~\cite{#1}}
\newcommand{\refref} [1] {ref.~\cite{#1}}

\newcommand{\refeq}  [1] {(\ref{#1})}
\newcommand{\reffig}[1]{Fig.~\ref{#1}}  
\newcommand{\reftab} [1] {table~\ref{#1}}

\newcommand{\beq}{\begin{equation}}

\newcommand{\eeq}{\end{equation}}
\newcommand{\ee}[1] {\label{#1} \end{equation}}
\newcommand{\bea}{\begin{eqnarray}}

\newcommand{\eea}{\end{eqnarray}}

\newcommand{\ie}{{i.e.}}            
\newcommand{\etal}{{\em et al.}}    

\newcommand{\statesp}{state space}

\newcommand{\pS}{\ensuremath{{\cal M}}}          
\newcommand{\ssp}{\ensuremath{x}}                
\newcommand{\slice}{slice}

\newcommand{\mslices}{method of slices}

\newcommand{\sliceBord}{slice border}

\newcommand{\slicePlane}{slice hyperplane}

\newcommand{\fFslice}{first Fourier mode slice}

\newcommand\period[1]{{\ensuremath{T_{#1}}}}         
\newcommand{\timeAver} [1]{\overline{#1}}
\newcommand{\pSRed}{\ensuremath{\hat{\cal M}}} 
\newcommand{\sspRed}{\ensuremath{\hat{\ssp}}}    

\newcommand{\slicep}{{\ensuremath{\sspRed'}}}   
\newcommand{\sliceTan}[1]{\ensuremath{t'_{#1}}}    
\newcommand{\groupTan}{\ensuremath{t}}    
\newcommand{\Group}{\ensuremath{G}}         
\newcommand{\LieEl}{\ensuremath{g}}  

\newcommand{\zeit}{\ensuremath{t}}  
\newcommand{\sspRSing}{\ensuremath{\sspRed^\ast}} 	
\newcommand{\shift}{\ensuremath{\ell}}

\newcommand{\velRel}{\ensuremath{c}}    
\newcommand{\phaseVel}{phase velocity}      

\newcommand{\sign}[1]{\omega_{#1}}

\newcommand{\eigExp}[1][]{
     \ifthenelse{\equal{#1}{}}{\ensuremath{\lambda}}{\ensuremath{\lambda^{(#1)}}}}
\newcommand{\eigRe}[1][]{
     \ifthenelse{\equal{#1}{}}{\ensuremath{\mu}}{\ensuremath{\mu^{(#1)}}}}
\newcommand{\eigIm}[1][]{
     \ifthenelse{\equal{#1}{}}{\ensuremath{\omega}}{\ensuremath{\omega^{(#1)}}}}

\newcommand{\po}{periodic orbit}

\newcommand{\rpo}{rela\-ti\-ve periodic orbit}

\newcommand{\eqva}{equi\-lib\-ria}

\newcommand{\reqv}{traveling wave}

\newcommand{\reqva}{traveling waves}
\newcommand{\Reqva}{Traveling waves}

\renewcommand{\vec}[1]{\mbox{\boldmath $#1$}} 

\begin{document}
\title{
Symmetry reduction in high dimensions, illustrated in a turbulent pipe
    }
\author{Ashley P. Willis }
\email[]{a.p.willis@shef.ac.uk}
\affiliation{
        School of Mathematics and Statistics,
        University of Sheffield, S3 7RH, U.K.
            }
\author{Kimberly Y. Short}
\email[]{kyshort@gatech.edu}
\author{Predrag Cvitanovi\'{c}}
\email[]{predrag@gatech.edu}
\affiliation{   Center for Nonlinear Science, School of Physics,
                Georgia Institute of Technology, Atlanta, GA
                30332-0430 }

\date{\today}

\begin{abstract}

Equilibrium solutions are believed to structure the pathways for ergodic
trajectories in a dynamical system. However, equilibria are atypical for
systems with continuous symmetries, i.e. for systems with homogeneous
spatial dimensions, whereas \textit{relative} equilibria (traveling waves)
are generic. In order to visualize the unstable manifolds of such
solutions, a practical symmetry reduction method is required that
converts relative equilibria into equilibria, and relative periodic
orbits into periodic orbits. In this article we extend the fixed Fourier
mode slice approach, previously  applied 1-dimensional PDEs, to a
spatially 3-dimensional  fluid flow, and show that is substantially more
effective than our previous approach to slicing. Application of this
method to a minimal flow unit pipe leads to the discovery of many
relative periodic orbits that appear to fill out the turbulent regions of
state space. We further demonstrate the value of this approach to
symmetry reduction through projections (projections only possible in the
symmetry-reduced space) that reveal the interrelations between these
relative periodic orbits and the ways in which they shape the geometry of
the turbulent attractor.
\end{abstract}


\maketitle

Chaotic dynamics can be interpreted as a trajectory in state space, where
each coordinate corresponds to a degree of freedom.  For
higher-dimensional systems it can be difficult to predict which
coordinate choices will provide the most instructive projections, given
that plots of these trajectories are limited to displaying two or three
dimensions at a time. To avoid clutter in the projection caused by
families of orbits related by translations or reflections,
symmetry-invariant measures such as spatial averages are often favored.
In practice, however, there are only so many quantities that may be
averaged and, in addition, information held in the spatial structure is
wiped out in the averaging process. Often such averaging results in a
largely uninformative projection of the dynamics.

The study of turbulence is one example where substantial progress has
recently been made by viewing the flow as a dynamical system, but now a
more informative means of projection is required to comprehend the way in
which the unstable manifolds of relative equilibria and other invariant
solutions shape the dynamics.  These invariant solutions correspond to
recurrent but unstable motions\rf{science04} that share some
characteristics with fully turbulent flows.
Experiments\rf{science04,DeSo14} and
simulations\rf{KeTu06,deLMeAvHo12} have identified transient visits to
spatiotemporal patterns that mimic {\reqv} solutions. Certain
low-dissipation \reqva\ of the Navier-Stokes equations have been shown
to be important in the transition to turbulence, where they lie in the
laminar-turbulent boundary, separating initial conditions that ultimately
relaminarize from those that develop into turbulence\rf{duguet07}.
Spatiotemporal flow patterns called `puffs' and `slugs' are observed
during the evolution to turbulence. Recently, spatially-localized
solutions representative of puffs have been discovered\rf{AvMeRoHo13} and
shown to be linked to spatially-periodic \reqva\ in minimal
domains\rf{ChWiKe14}.
As \reqva\ are steady in their respective co-moving frames,
they are {\em relative} equilibria,
solutions that do not exhibit temporal shape-changing dynamics.
Their unstable manifolds, however, mold the surrounding \statesp, carving
pathways for relative periodic orbits,
invariant orbits embedded in turbulence whose temporal evolution captures
dynamics of ergodic trajectories that shadow them. A detailed
understanding of these recurrent motions is crucial if one is to
systematically describe the repertoire of all turbulent motions.
With the removal of spatial translations, which obscure visualizations of
the dynamics, a far greater number of projections of chaotic trajectories
is possible. In this article, we show that visualizations of the symmetry
reduced dynamics can help us understand relationships between distinct
families of periodic orbits and traveling wave solutions, which in turn lends
support to the dynamical systems interpretation that {\em relative}
periodic orbits form the backbone of turbulence in pipe.

Our approach is dynamical: writing the Navier--Stokes equations as
$\dot{\vec{u}} = \vec{v}(\vec{u})$, the fluid state $\vec{u}$ at a
particular moment in time is represented by a single point in \statesp\
$\pS$\rf{GHCW07}; turbulent flow is represented by an ergodic trajectory
that wanders between accessible states in $\pS$\rf{Hopf48}. Essential to
this analysis is that any two physically equivalent states be identified
as a single state: a symmetry-reduced \statesp\ $\pSRed=\pS/\Group$ is
formed by contracting the volume of \statesp\ representing states that
are identical except for a symmetry transformation to a single point
$\hat{\vec{u}}$. Only after a symmetry reduction are the relationships
between physically distinct states revealed.
In this article symmetry reduction is implemented with
an extension of the `\fFslice' method\rf{BudCvi14}, a variant of the
\mslices\rf{CartanMF}.
The \mslices\ separates
coordinates into phases along symmetry directions (`fibers', `group
orbits' that parametrize families of physically-equivalent dynamical
states) from the remaining coordinates of the symmetry-reduced \statesp\
$\pSRed$.  The latter capture the dynamical degrees of freedom---those
associated with structural changes of the flow.

\setlength{\unitlength}{0.45\textwidth}
  \begin{figure}
  \begin{center}
\includegraphics[width=\unitlength]{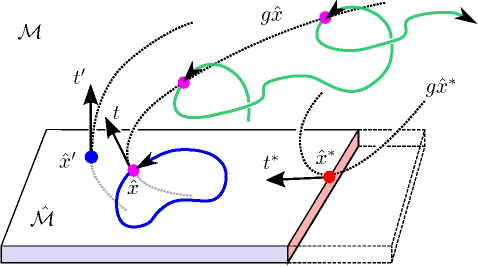}
\end{center}
   \caption{ \label{fig:mslices}
     (Color online)
     Schematic of symmetry reduction by the \mslices. The blue point is
     the template $\slicep$. Group orbits are marked by dotted curves, so
     that all pink points are equivalent to $\sspRed$ up to a shift.  The
     \rpo\ (green) in the $d$-dimensional full \statesp\ $\pS$  closes
     into a \po\ (blue) in the slice $\pSRed=\pS/\Group$, a
     $(d\!-\!1)$-dimensional slab transverse to the template group
     tangent $\sliceTan{}$. A typical group orbit crosses the
     {\slicePlane} transversally, with a non-orthogonal group tangent
     $\groupTan=\groupTan(\sspRed)$. A {\slicePlane} is almost never a
     global \slice; it is valid up to the {\sliceBord}, a
     $(d\!-\!2)$-dimensional hypersurface (red) of points $\sspRSing$
     whose group orbits graze the slice, \ie\ points whose tangents
     $\groupTan^*=\groupTan(\sspRSing)$ lie in $\pSRed$. Beyond the
     {\sliceBord} (dashed `chunk'), group orbits do not cross the slice
     hyperplane locally.
   }
 \end{figure}

The Navier-Stokes equations are invariant under translations, rotations,
and inversions about the origin, and the application of any of these
symmetry operations to a state $\vec{u}(\vec{x})$ results in another
dynamically equivalent state. The boundary conditions for pipe flow
restrict symmetries to translations along the axial and azimuthal
directions, and reflections in the azimuthal direction. In the
computations presented here,axial periodicity is assumed so
that the symmetry group of the system is $O(2)_\theta\times SO(2)_z$. In
order to illustrate the key ideas, we constrain azimuthal shifts, and
focus on the family of streamwise translational shifts $\{\LieEl\}$
parametrized by a single continuous phase parameter $\shift$,
\[
  (\LieEl(\shift)\, \vec{u})(z) = \vec{u}(z-\shift) \, .
\]
If periodic axial symmetry is assumed, application of $\LieEl$ gives a
closed curve family of dynamically equivalent states  --- topologically a
circle, called a \textit{group orbit} --- in \statesp\ $\pS$. Were
azimuthal (`spanwise') shifts included, equivalent states would lie on a
2-torus.

Symmetry reduction simplifies the \statesp\ by reducing each set of
dynamically equivalent states to a unique point $\hat{\vec{u}}$. The
\textit{\mslices} achieves this with the aid of a fixed \textit{template}
state $\vec{u}'$ (see \reffig{fig:mslices}). A shift is applied so that
the symmetry-reduced state
$\hat{\vec{u}}=\LieEl(-\shift) \vec{u}$
lies within the hyperplane orthogonal to
$\vec{t}'=\lim_{\shift\to0}\,(\LieEl(\shift) \vec{u}'- \vec{u}')/\shift$,
the tangent to the template $\vec{u}'$ in the direction of the shift. For
a time-dependent flow, one determines $\shift=\shift(t)$ by chosing
$\hat{\vec{u}}$ to be the point on the group orbit of $\vec{u}$ closest
to the template, $\langle \hat{\vec{u}}-\vec{u}'|\vec{t}'\rangle=0$ in a
given norm. In this work we use the L2 or `energy' norm
$E = \langle \vec{u}|\vec{u}\rangle/2 = \int \vec{u}^2/2\,dV$.

As \reqva\ drift downstream without changing their spatial structure, the
family of \reqv\ states $\vec{u}(t)$ is dynamically equivalent (lies on
the same group orbit $\LieEl(\shift)\, \vec{u}$) and may be represented
by a single state $\hat{\vec{u}}_q$. Thus all \reqva\ are simultaneously
reduced to \eqva\
in the \slice, irrespective of their individual phase velocities,
a powerful property of the \mslices. Furthermore, all \rpo s $p$, flow
patterns each of which recurs after a different time period $\period{p}$,
shifted downstream by a different $\shift_p$, close into \po s in the
{\slicePlane}.

Dynamics within the slice is given by
\bea
\dot{\hat{\vec{u}}} &=& \vec{v}(\hat{\vec{u}})
- \dot{\shift}(\hat{\vec{u}}) \, \vec{t}(\hat{\vec{u}}) \, , \\
\label{eq:slice} \dot{\shift} (\hat{\vec{u}}) &=& \langle \vec{v}
(\hat{\vec{u}}) | \vec{t}'\rangle \, / \, \langle \vec{t}(\hat{\vec{u}})
| \vec{t}'\rangle
\, ,
\label{eq:reconst}
\eea
where the expression for
the  {\phaseVel} $\dot{\shift}$ is known as the {\em reconstruction
equation}\rf{rowley_reconstruction_2000}. No dynamical information is
lost and we may return to the full space by integrating
\refeq{eq:reconst}. In contrast to a Poincar\'e section, where
trajectories pierce the section hyperplane, time evolution traces out a
continuous trajectory within the slice. In principle, the choice of
template is arbitrary; in practice, some templates are preferable to
others.
While one is concerned with the dynamics within the slice
$\hat{\vec{u}}(t)$, in practice it may be simpler to record $\shift(t)$
and to post-process, or to process on the side, visualizations within the
slice---slicing is much cheaper to perform than gathering $\vec{u}(t)$
from simulation or laboratory experiment.

The enduring difficulty with symmetry reduction is in determining a
unique shift $\shift$ for a given state $\vec{u}$, while avoiding
discontinuities in $\shift(t)$ that arise when multiple `best fit'
candidates $\hat{\vec{u}}=\LieEl(-\shift)\vec{u}$ to the template
$\vec{u}'$ occur. A singularity arises if the group orbit $g\vec{u}$
grazes the slice hyperplane (\reffig{fig:mslices}). At the instant this
occurs, the tangents to the fluid state $\hat{\vec{u}}$ and the template
$\vec{u}'$ are orthogonal, and there is a division by zero in the
reconstruction equation \refeq{eq:reconst}.
In \refref{ACHKW11} it was
shown that the hyperplanes defined by multiple templates could be used to
tile a slice,
but while switching may permit the symmetry reduction of
longer trajectories, it is often not possible to both switch
templates before a {\sliceBord} is reached and to simultaneously
maintain continuity in $\shift$.  Furthermore, it is uncertain when
to switch back to the first template, in order to produce a unique
symmetry-reduced state. Our aim in this article is to avoid such
difficulties through the use of a single template with distant
\sliceBord s.
The approach of Budanur \etal\rf{BudCvi14} for the case of one translational
spatial dimension fixes the phase of a single Fourier coefficient.
This `Fourier' slice is a special case within the slicing framework,
with the effect of extreme smoothing of the group orbit.
Here the approach is extended to a spatially 3-dimensional case, that of
turbulent pipe flow.

\begin{figure}
  \centering
\includegraphics[width=0.18\textwidth]{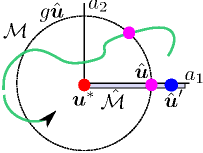}%
~~\includegraphics[width=0.22\textwidth]{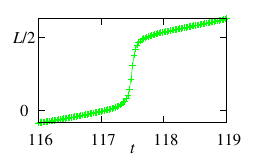} \\[-2pt]
  \includegraphics[width=0.45\textwidth]{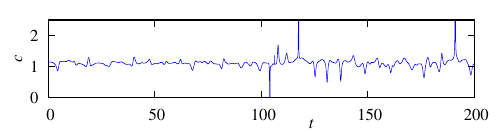}
   \caption{ \label{fig:6341sz}
({\it top left}) Schematic of the {\fFslice}, with $a_1,\,a_2$ defined in
the text. In this projection the {\sliceBord} is a zero-measure `point'
at the origin. ({\it bottom}) For a generic ergodic trajectory the
{\phaseVel} $c=\dot{\shift}(t)$ appears to encounter singularities
whenever it approaches the {\sliceBord}, which, however, is never
reached\rf{BudCvi14}. Closer inspection reveals a  rapid but continuous
change in the shift  ({\it top right}) by $\approx L/2$ in the
$\shift(\zeit)-\timeAver{\velRel}\zeit$ Galilean frame, moving
(for our parameter values) at
$\timeAver{\velRel}=1.1092$. These apparent jumps are well resolved: each
`+' corresponds to 10 time integration steps.
   }
 \end{figure}

For the case of a scalar field defined on one spatial
dimension\rf{BudCvi14} there is a unique Fourier coefficient appropriate
for determining the symmetry reduction. Here, for the 3-dimensional
turbulent flow, there are three components of velocity with a spatial
discretization for each, and it is not obvious which coefficients to fix
in order to define an effective symmetry-reducing slice.
In this paper we construct a template
$\vec{u}'(r,\theta,z)=\vec{u}_c\cos(\alpha z)
+ \vec{u}_s\sin(\alpha z)$, where
$\vec{u}_c(r,\theta)=\int_0^L
\tilde{\vec{u}}\cos(\alpha z) \, \mathrm{d}z$,
$\vec{u}_s(r,\theta)=\int_0^L
\tilde{\vec{u}}\sin(\alpha z) \, \mathrm{d}z$,
and $L=2\pi/\alpha$, for some chosen state $\tilde{\vec{u}}$.
This corresponds to (all of) the first coefficients in the streamwise
Fourier expansion for $\tilde{\vec{u}}$.
Arbitrary states $\vec{u}$ may then be projected onto a plane via
$a_1=\langle\vec{u}|\vec{u}'\rangle$ and
$a_2=\langle\vec{u}\,|\,\LieEl(L/4)\,\vec{u}'\rangle$, respectively (see
\reffig{fig:6341sz}).
In this projection, the group orbit $\LieEl\vec{u}$
of any state is a circle
centered on the origin,
and the polar angle $\theta$ for the point $(a_1,a_2)$
corresponds to a unique shift $\shift=\theta(L/2\pi)$. The symmetry
reduced state $\hat{\vec{u}}=\LieEl(-l)\,\vec{u}$ is the
closest point on its group orbit to the template
$\vec{u}'$.
The slice is projected onto the positive $a_1$-axis in this projection.

Note that the approach is independent of discretization, and does not
actually require a Fourier decomposition. Note also that the
inner-product gathers information from the full velocity field.

As group orbits are circles crossing perpendicular to the $a_1$-axis
in this projection, $\langle \vec{t}(\hat{\vec{u}})|\vec{t}'\rangle$ in
\refeq{eq:reconst} can only be zero if the circle shrinks to a point at
the origin. This requires that both inner products
$\langle\vec{u}|\vec{u}'\rangle$ and
$\langle\vec{u}\,|\,\LieEl(L/4)\,\vec{u}'\rangle$ are zero at the same
time, which has vanishing probability. While we thus avoid the
{\sliceBord}, there is a rapid change in $\theta$ by $\approx\pi$ (in
$\shift$ units by $\approx L/2$) whenever the trajectory
$(a_1,a_2)(\zeit)$ sweeps past the origin, see the inset to
\reffig{fig:6341sz}.  Rapid phase shifts notwithstanding, this choice
of template has made possible the discovery and analysis of the many \rpo
s discussed below.

\begin{table}
   \begin{tabular}{lcccrrrc}  
   \hline
       & $\timeAver{D}$
       & $\timeAver{\velRel}$
				  & \#
                                   & $D_{KY}$
                                   & ~$\eigRe[max] $
                                   & $ \sign{}$ or $\theta$ \\
                                      \hline

   \TWRUN{N4L/1.38}{6491}    & 1.380 & 1.238 &   3  & 6.97 &  0.1809  & 0\\
   \TWRUN{2.03}{6494} & 2.039  & 1.091  &   7   &  15.21 &  0.1159 & 0\\
   \TWRUN{1.97}{6472}   & 1.968  & 1.104  &  9  & 20.01 &  0.1549  & 0.259\\
   \TWRUN{2.04}{8014} & 2.041  & 1.095  &   8   & 20.04 &  0.1608 & 0\\
   \TWRUN{N4U/3.28}{6482} &  3.279 & 1.051 & 30  & 73.67   &  0.9932 & 3.136
   \\
   \hline
\RPORUN{6.66}{6350}  &  1.806 &  1.122  &  3 &  7.99 &  0.0535 & 1.690   \\
\RPORUN{27.30}{6655} &  1.815 &  1.127  &  4 &  8.98 &  0.0678 & 0.961   \\
\RPORUN{13.19}{6652} &  1.839 &  1.119  &  5 &  9.68 &  0.0581 & 2.038   \\
\RPORUN{20.43}{6465} &  1.809 &  1.130  &  5 & 11.03 &  0.0771 &   +1    \\
\RPORUN{4.95}{8100}  &  2.015 &  1.090  &  3 & 11.54 &  0.1509 & 1.643   \\
\RPORUN{7.72}{8058}  &  1.708 &  1.141  &  5 & 11.62 &  0.0983 &    +1   \\
\RPORUN{15.46}{6607} &  1.781 &  1.027  &  7 & 12.69 &  0.1162 &    +1   \\
\RPORUN{9.74}{8027}  &  2.050 &  1.088  &  7 & 12.87 &  0.1873 &    -1   \\
\RPORUN{23.36}{8087} &  1.980 &  1.113  &  6 & 13.37 &  0.1011 & 1.251   \\ 
\RPORUN{7.42}{8024}  &  1.838 &  1.111  &  6 & 13.89 &  0.1195 & 0.388   \\
\RPORUN{17.46}{8142} &  1.917 &  1.122  &  6 & 14.67 &  0.0841 & 0.196   \\ 
\RPORUN{14.05}{6477} &  1.902 &  1.109  &  7 & 14.75 &  0.1403 &    -1   \\

  ergodic & 1.956 & 1.109 \\ \hline
   \end{tabular}
   \caption{\label{tab:RPOdata_PRL}
A subset of {\reqva} and \rpo s of the lowest Kaplan-Yorke
dimension\rf{FKYY83}, out of respectively 10 and 29 extracted so far and
plotted in \reffig{fig:PCAplot}. {\Reqva} are labeled by their
dissipation rate, and \rpo s are labeled by their period \period{}. Listed
are mean dissipation ${\timeAver{D}}$, mean down-stream {\phaseVel}
$\timeAver{\velRel}$, the number of unstable eigen-directions (two per
each complex pair), Kaplan-Yorke dimension $D_{KY}$, the real part of the
largest stability eigenvalue/Floquet exponent $\eigRe[max]$, and either
the corresponding imaginary part $\eigIm[max]$ for \reqva, or the phase
$\theta$ of the complex Floquet multiplier for \rpo s, or its sign, if
real: -1 indicates inverse hyperbolic.
   }
\end{table}

`Minimal flow units'\rf{JM91}, which capture much of the statistical
properties of turbulence, have been invaluable in analyzing  fundamental
self-sustaining processes\rf{HaKiWa95}.
Here, the fixed-flux Reynolds number for all calculations is
$Re=DU/\nu=2500$, where lengths are non-dimensionalized by diameter $D$
and velocities are normalized by the mean axial speed $U$. The minimal
flow unit is in the $m=4$ rotational subspace, such that
$(r,\theta,z)
\in [0,\frac{1}{2}]\times[0,\frac{\pi}{2}]\times[0,\frac{\pi}{1.7}]$.
The size of the domain is more usefully measured in terms of wall units,
$\nu/u_\tau$, where
$u_\tau^2=-\nu\left.(\partial_ru_z)\right|_\mathrm{wall}$,
which allows comparison with flow units used in other geometries.
In these units,
the domain is of size $\Omega^+\approx[100,160,370]$
in the wall-normal, spanwise and streamwise dimensions, respectively. Our
flow unit compares favorably with the minimal flow units for channel
flow\rf{JM91} $\Omega^+\approx[\,>\hspace{-2pt}40,100,250-350]$ and
Couette flow\rf{HaKiWa95} $\Omega^+\approx[68,128,190]$. Recurrent flows
have been identified in \refref{GHCW07} for a box of size
$\Omega^+\approx[68,86,190]$.
Our domain is sufficiently large to reproduce
$Re_\tau=(D/2)u_\tau/\nu=100\pm1$ to within 10\% of its value
in the infinite domain.  The mean wall friction for turbulent flow is
approximately 100\% greater than that for laminar flow at this
flow rate.

\setlength{\unitlength}{.48\textwidth}
   \begin{figure}
   \centering
      \includegraphics[width=\unitlength]{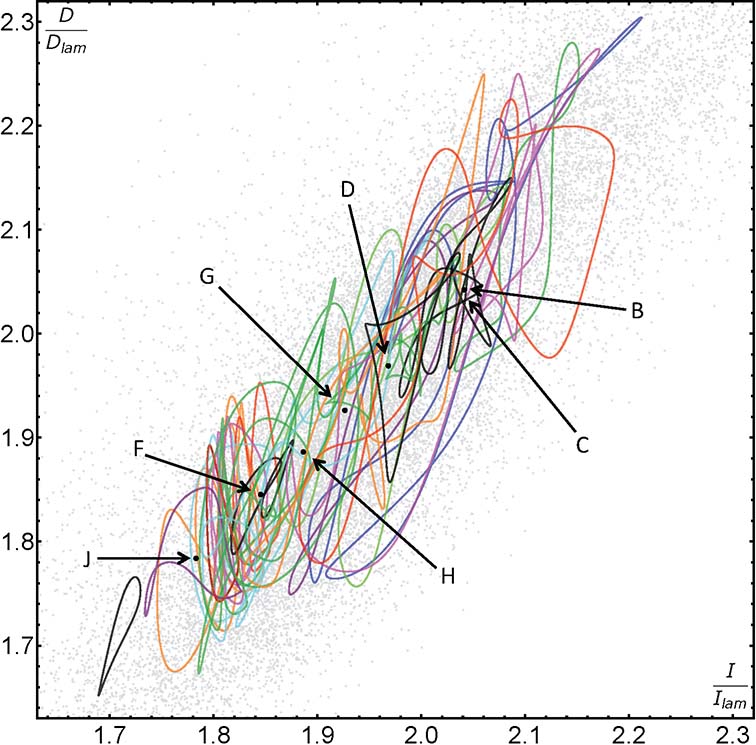}
   \caption{ \label{fig:idplot}
     (Color online)
Projection of 32 \rpo s and \reqva\
using symmetry-invariant coordinates,
$I/I_{lam}$, $D/D_{lam}$ where $I_{lam}=D_{lam}$ are
energy rates for the laminar flow.
All discovered \reqva\ are included:
(B) $\TWRUN{2.04}{6480}$,
(C) $\TWRUN{2.03}{6494}$,
(D) $\TWRUN{1.97}{6472}$,
(G) $\TWRUN{1.93}{6495}$,
(H) $\TWRUN{1.89}{6491}$,
(F) $\TWRUN{1.85}{6416}$ and
(J) $\TWRUN{1.78}{6459}$,
except for
$\TWRUN{N4L/1.38}{}$, $\TWRUN{N4U/3.28}{6482}$ and $\TWRUN{1.57}{}$,
which lie far outside the ergodic cloud (grey dots).
        }
 \end{figure}

A Newton-Krylov scheme is used to search for \rpo s. Initial guesses are
taken from near recurrences of ergodic trajectories\rf{GHCW07} within the
symmetry-reduced \statesp.  This preferentially identifies structures
embedded in regions of high natural measure (regions most frequented by
ergodic trajectories), with isolated \reqva\ and \rpo s that sit in the
less frequented reaches of \statesp\ less likely to be found. Our
searches have so far identified 10 \reqva\ and 32 \rpo s. An abbreviated
summary of data is given in \reftab{tab:RPOdata_PRL}; the complete data
set is available online at \texttt{Openpipeflow.org}, along with the open
source code used to calculate these orbits.

Visualizations of high-dimensional \statesp\ trajectories are necessarily
projections onto two or three dimensions. A common choice is to monitor
the flow in terms of the rate of energy dissipation
$D=\rho\nu\int\vec{u}\cdot{\bf\nabla}^2\vec{u}\,dV$ and 
the external input power required to maintain constant flux
$I=Q\,\Delta p$, where 
$Q=\int\vec{u}\cdot\vec{dS}$ is the flux at any cross-section and
$\Delta p$ and is the pressure drop over the length of the pipe.
As the time-averages of $I$ and $D$ are
necessarily equal, \reqva\ and orbits, which may be well-separated in
\statesp, are contracted onto or near the $I=D$ line, a drawback of the
2-dimensional $(I,D)$ projection.
\reffig{fig:idplot} shows that the orbits appear to overlap with
the ergodic region, but reveals
little of the relationships between solutions; we use $D$ values only to
distinguish \reqva\ solutions listed in \reftab{tab:RPOdata_PRL}.

\setlength{\unitlength}{.48\textwidth}
   \begin{figure}
   \centering
      \includegraphics[width=\unitlength]{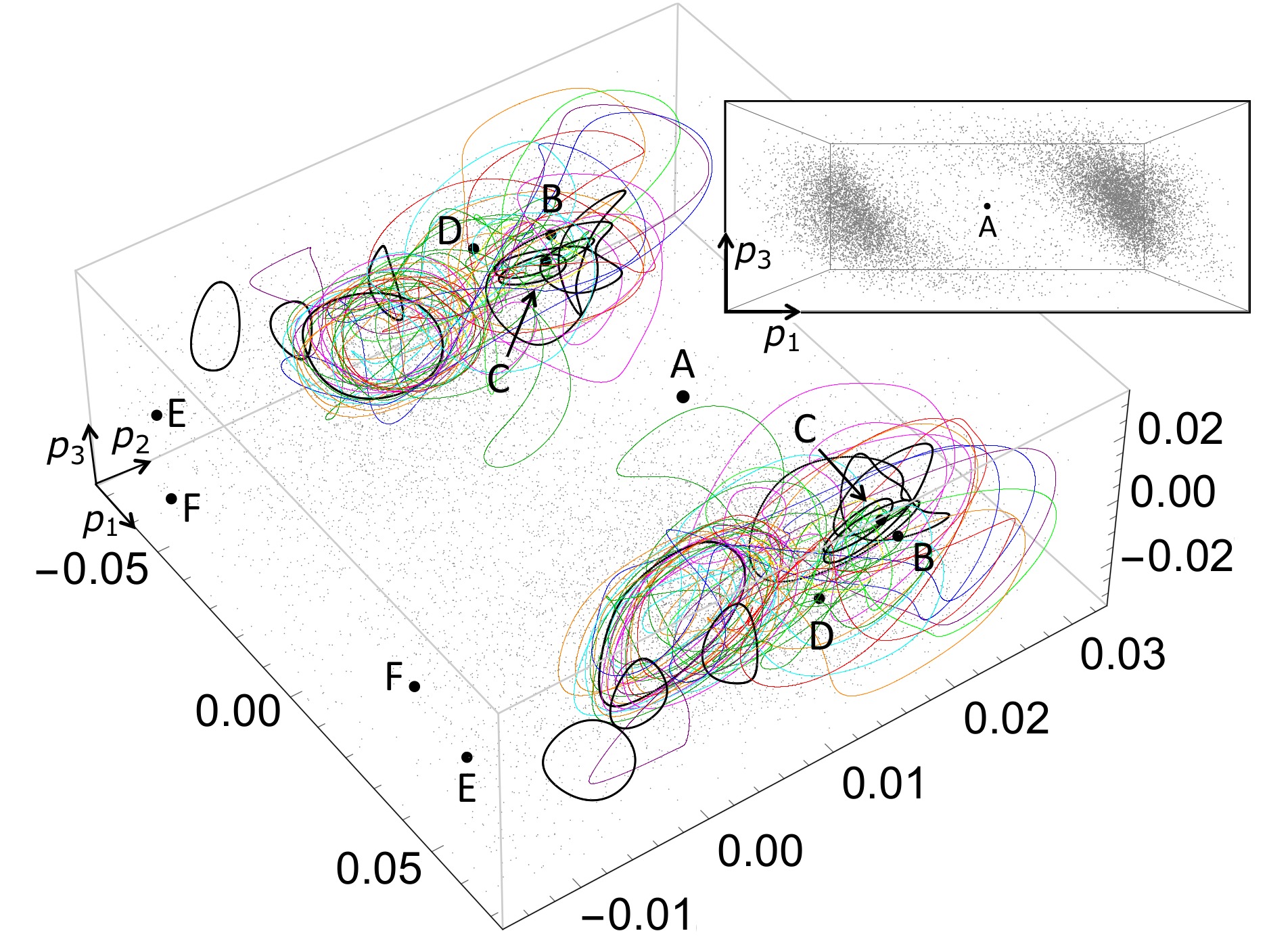}
   \caption{ \label{fig:PCAplot}
     (Color online)
Projection of the symmetry-reduced infinite-dimensional {\statesp}
onto the first 3 PCA principal axes, computed from the L2-norm average
over the natural measure (the gray `cloud') in the slice.
  32 \rpo s, and
  a subset of the 7 shortest \rpo s, together with \reqva\
(A) $\TWRUN{N4U/3.28}{6482}$,
(B) $\TWRUN{2.04}{8014}$,
(C) $\TWRUN{2.03}{6494}$,
(D) $\TWRUN{1.97}{6472}$,
(E) $\TWRUN{1.98}{6491}$,
(F) $\TWRUN{1.85}{6416}$.
While
$\TWRUN{1.93}{6495}$
appears to lie in the very center of 
the $(I,D)$ projection \reffig{fig:idplot},
it is revealed
in this \statesp\ projection 
to lie far from the ergodic cloud, outside the box plotted,
as are (E) $\TWRUN{N4L/1.38}{6491}$
and (F) $\TWRUN{1.85}{6416}$. Due to a `rotate-and-reflect' symmetry,
each solution appears twice, with the exception of (A)
$\TWRUN{N4U/3.28}{6482}$ (and the far-away $\TWRUN{N4L/1.38}{6491}$), which
belong to the `rotate-and-reflect' invariant subspace. 
Our \rpo s capture
the regions of high natural measure very well.
The symmetry-invariant subspace has a
strong repulsive influence, separating the natural measure into two
weakly communicating regions.  The inset shows the ergodic cloud from 
another perspective.
           }
 \end{figure}

In the symmetry-reduced \statesp\ it is possible to construct coordinates
that are intrinsic to the flow itself, using spatial information that
would otherwise be smeared out by translational shifts. To obtain a
global portrait of the turbulent set, \reffig{fig:PCAplot}, we project
solutions onto the three largest principal components $\hat{\vec{e}}_i$
obtained from a PCA of $N$=2000 independent
$\hat{\vec{u}}'_i=\hat{\vec{u}}_i-\bar{\hat{\vec{u}}}$, where
$\bar{\hat{\vec{u}}}$ is the mean of the data, using the SVD method 
(on average the square of the projection
$p_i=\langle\hat{\vec{u}}'(t)|\hat{\vec{e}}_i\rangle$ equals the
$i^\mathrm{th}$ singular value of the correlation matrix 
$R_{ij}=\frac{1}{N-1}\langle\hat{\vec{u}}'_i|\hat{\vec{u}}'_j\rangle$).

The lower / upper branch pair \TWRUN{N4L/1.38}{6491} /
\TWRUN{N4U/3.28}{6482} were obtained by continuation from a smaller
`minimal flow unit'\rf{ACHKW11}. In \reftab{tab:RPOdata_PRL} and in the
$(I,D)$-projection \reffig{fig:idplot}
the upper branch \reqv\ \TWRUN{N4U/3.28}{6482} appears
to be far removed from turbulence, unlikely to exert influence. The PCA
projection of the symmetry-reduced \statesp, however, reveals the strong
repelling influence of \TWRUN{N4U/3.28}{6482} whose 30-dimensional
unstable manifold acts as a barrier to the dynamics, cleaving the natural
measure into two `clouds', forcing a trajectory to hover around one
neighborhood until it finds a path to the other, bypassing
\TWRUN{N4U/3.28}{6482}. The two ergodic `clouds' are related by the
`rotate-and-reflect' symmetry ($\pi/4$ rotation), under which
\TWRUN{N4U/3.28}{6482} is invariant  (for symmetries of pipe flow see
\refref{ACHKW11}).

The symmetry-reduced \statesp\ projections reveal sets of \rpo s with
qualitatively similar dynamics. The short-period orbits are well spread
over the dense regions of natural measure, and the long \rpo s in (a)
appear to `shadow' short orbits in (b), but also exhibit extended
excursions that fill out {\statesp}. While sets of \rpo s often share
comparable dissipation rates and Floquet exponents
(\reftab{tab:RPOdata_PRL} and \texttt{Openpipeflow.org} data sets), it is
the \statesp\ projections that are essential to establishing genuine
relationships.

In summary, we have shown that symmetry reduction
can be applied to a dynamical system of very high dimensions,
here turbulent pipe flow.
An appropriately constructed template renders the \mslices\ substantially
more effective for
projecting the dynamics and for
Newton searches for invariant solutions.
The method is general and can be applied to any dynamical system with
continuous translational or rotational symmetry. Projections of the
symmetry-reduced space reveal fundamental properties of the dynamics not
evident prior to symmetry reduction. In the application at hand, to a
turbulent pipe flow, the method has enabled us to identify for the first
time a large set of \rpo s embedded in turbulence, and to demonstrate
that the key invariant solutions strongly influence turbulent dynamics.
To follow this demonstration of the power of symmetry reduction,
work is now underway to determine the relationship between \rpo s\rf{WFSBC15}.
Analysis of their
unstable manifolds are expected to reveal the intimate links between
\reqva\ and \rpo s, allowing for explicit construction of the invariant
skeleton that gives shape to the strange attractor explored by
turbulence.

\begin{acknowledgments}

We are indebted to
M.~Farazmand,
N.~B.~Budanur,
J.F.~Gibson,
X.~Ding,
F.~Fedele,
E.~Siminos,
M.~Avila,
B.~Hof,
and
R.~R.~Kerswell
for many stimulating discussions.
A.\,P.\,W.\ is supported by the EPSRC under grant EP/K03636X/1.
K.\,Y.\,S.\ was supported by the National Science Foundation Graduate
Research Fellowship under Grant NSF~DGE-0707424.
P.\,C.\ thanks the family of late G.~Robinson,~Jr.\
and
NSF~DMS-1211827 for support.

\end{acknowledgments}

\bibliographystyle{apsrev4-1}
\bibliography{../bibtex/pipes}

\end{document}